\documentstyle[aps]{revtex}
\begin{document}
\draft 
\preprint{gr-qc/yymmxxx} 
\title{Black hole entropy : quantum vs thermal fluctuations } 
\author{Ashok Chatterjee\footnote{email:
ashok@theory.saha.ernet.in} and Parthasarathi Majumdar\footnote{On
deputation from the Institute of Mathematical Sciences, Chennai 600
113, India; email: partha@theory.saha.ernet.in}}
\address{Theory Group, Saha Institute of Nuclear Physics, Kolkata 700
064, India.}  
\maketitle
\begin{abstract}
The relation between logarithmic corrections to the area
law for black hole entropy, due to thermal fluctuations
around an equilibrium canonical ensemble, and those
originating from quantum spacetime fluctuations within a
microcanonical framework, is explored for three and four
dimensional asymptotically anti-de-Sitter black holes. For
the BTZ black hole, the two logarithmic corrections are
seen to precisely cancel each other, while for four
dimensional adS-Schwarzschild black holes a partial
cancellation is obtained. We discuss the possibility
of extending the analysis to asymptotically flat black holes.
\end{abstract}

\section{Introduction}

Non-perturbative canonical Quantum General Relativity (QGR) \cite{aa1}
reveals an appealing picture \cite{abk} of microstates underlying the
area law \cite{bek,haw} for black hole entropy. One begins with a
classical connection formulation of a black hole horizon where the
usual event horizon (appropriate to a stationary situation) is
replaced by an {\it isolated} horizon, defined entirely by boundary
conditions locally, i.e., without reference to asymptotic null
infinity ${\cal I}_{\pm}$.  These boundary conditions lead uniquely to
a Chern Simons theory `living' on the horizon, with the bulk geometry
(the pullback of the densitised triad to the two-sphere foliation of
the horizon) playing the role of source currents for that
theory. Quantization, using the spin network formalism of canonical
QGR leads to a description of spacetime fluctuations on the horizon in
terms of states of this Chern Simons theory. Links of the bulk spin
network puncturing the horizon provide pointlike sources for this
Chern Simons theory. Black hole entropy, defined in terms of the
degeneracy of the boundary Chern Simons states, can be evaluated
either by solving the theory directly \cite{abk}, or by relating it to
the dimensionality of the Hilbert space of the two dimensional
boundary Wess-Zumino-Witten theory \cite{km}. This dimensionality can
be evaluated using standard techniques of two dimensional conformal
field theory. In both cases, the area law ensues, with a specific
choice of the Barbero-Immirzi parameter needed to obtain Hawking's
normalization of it. But the QGR formulation does more than simply
reproduce a law that has been anticipated decades ago on the basis of
semiclassical arguments; it provides an entire infinite series of
quantum corrections to the area law, in decreasing powers of horizon
area for macroscopic black holes with large horizons. Each term in the
series has a fixed, finite coefficient, completely calculable in
principle. The leading correction is logarithmic in the area, with the
coefficient -3/2 \cite{km2}, \cite{dkm}.

From a thermodynamic perspective, the `log area'
corrections can be thought of as `finite size' corrections
to the thermodynamic limit of large areas. The isolated
horizon boundary conditions require that nothing crosses
the horizon, although radiation may exist arbitrarily close
to it. The resulting constancy of the horizon area is
therefore a restriction of crucial importance, implying
that a {\it microcanonical} framework has been employed to
calculate the entropy. \footnote{For the Schwarzschild
black hole, this translates semiclassically into a
restriction on the black hole mass, related to the ADM
energy through addition of the radiant energy outside the
horizon.} This entropy is thus rightly called the {\it
microcanonical} entropy. The horizon area never undergoes
thermal fluctuations, because by fixing the area
(energy) of the black hole horizon,
such fluctuations have been eliminated. However, this does
appear to be overly restrictive in physical terms; even the
slightest amount of matter or radiation crossing the
horizon will violate it. In other words, a {\it canonical},
rather than microcanonical, ensemble seems to be more
suitable on physical grounds, since it allows thermal
fluctuations of various quantities like energy (area).

The effect of such thermal fluctuations on equilibrium
entropy has been investigated for asymptotically
anti-de-Sitter black holes \cite{dmb} using a canonical
ensemble\cite{hp}, and found to produce leading corrections
logarithmic in the area (for large area) with calculable
coefficients. These corrections were computed for
adS-Schwarzschild black holes of arbitrary dimension
including and beyond four, and also for the three
dimensional BTZ black hole. Links were also established
with an underlying conformal field theory structure.

The purpose of this paper is to clarify the precise relationship
between the logarithmic corrections to the area law originating from
quantum fluctuations of spacetime geometry, and those due to thermal
fluctuations in a canonical ensemble of black holes where the area is
no longer constrained to be a constant. This is done by deriving a
relation, using standard equilibrium statistical mechanics, between
the canonical entropy (including fluctuations around thermal
equilibrium), and the microcanonical entropy which equals the
logarithm of the density of states. The distinct roles played by
finite size corrections and thermal fluctuation corrections to black
hole entropy are thus made explicit.

In Section II, major tenets of the canonical QGR
computation of the microcanonical entropy are reviewed for
general four dimensional non-rotating black holes. In
Section III, the derivation of the relation between
canonical and microcanonical entropy for general
equilibrium statistical mechanical systems is presented.
The effect of thermal fluctuations is included in the
derivation. In subsection B of the same section, the
relation is employed to compute the canonical entropy of
asymptotically adS black holes in three and four
dimensions, using results of earlier computation of the
microcanonical entropy. The analysis thus far presented
does not extend to the more commonly studied asymptotically
flat black holes, which seem to have some kind of
instability, variously described in the literature as
`negative specific heat', `exponentially exploding density
of states' and so on. In Section V we show how this
problem manifests in a straightforward manner through the
relation derived in Section III. We discuss how the problem
may perhaps be obviated within canonical QGR.

\section{Microcanonical entropy}
\subsection{Classical Aspects}
This calculation \cite{abk} employs a classical Hamiltonian formulation of
general relativity, in terms of a canonical pair consisting
of the real $SU(2)$ connection $^{\gamma}A_a^i \equiv \Gamma_a^i + \gamma
K_a^i$ and the rescaled densitized triad $\gamma^{-1} E^{ai}$, with
$E^{ai} \equiv \det e e^{ai}$, where $e^{ai}$ is the triad on a chosen
spatial slice $M$. Here, $\Gamma_a^i \equiv \frac12 q_{ab} \epsilon^i_{jk}
\Gamma^{bjk}$, with $\Gamma^{bjk}$ being the pullback of the
Levi-Civita spin connection to the
spatial slice under consideration, and $q_{ab}$ is the 3-metric on the
slice; the extrinsic curvature $K_a^i$ is defined as $K_a^i \equiv
q_{ab}\Gamma^{b0i} $; $\gamma$, the Barbero-Immirzi parameter
\cite{imm} is a real positive parameter. Four dimensional local
Lorentz invariance has been partially gauge fixed to the `time gauge'
$e_a^0=-n_a$ where $n_a$ is the normal to the spatial slice. This
choice leaves the residual gauge group to be $SU(2)$. It is convenient
to introduce the quantity $^{\gamma}\Sigma_{ab}^{ij}
\equiv \gamma^{-1}e_{[a} ^ie_{b]}^i$, in terms of which the symplectic
two-form of general relativity can be expressed as 
\begin{eqnarray}
\Omega~=~{1 \over 8\pi G}~\int_M Tr [\delta ^{\gamma} \Sigma \wedge
\delta ^{\gamma} A'~-~\delta ^{\gamma} \Sigma' \wedge \delta ^{\gamma}A].
\label{sym}
\end{eqnarray}
The expression (\ref{sym}) of course is subject to modification by
boundary terms arising from the presence of boundaries of
spacetime. The black hole horizon, assumed to have the topology $S^2
\otimes R$, is intersected by $M$ in a two-sphere which thus plays the
role of an inner boundary.

Rather than using the notion of event horizon appropriate to
stationary situations studied in earlier literature \cite{bch}, we
adopt here the concept of `isolated' horizon \cite{abf}. This has the
advantage of being characterized completely locally, without requiring
a global timelike Killing vector field. The
characterization, for non-rotating situations, involves a null surface $H$
with topology as assumed above, with preferred foliation by two-
spheres and ruling by lines transverse to the spheres. $l^a$ and $n_a$
are null vector fields satisfying $l^an_a=-1$ on the isolated
horizon. $l^a$ is a tangent vector to the horizon, which is assumed to
be geodesic, twist-free, divergenceless and most importantly, {\it
non-expanding}. The Raychaudhuri equation is then used to prove that
it is also free of shear. Similarly, the null normal one-form field
$n_a$ is assumed to be shear- and twist-free, and have negative
spherical expansion. Finally, while stationarity is not a part of the
characterization of an isolated horizon, the vector direction field
$l^a$ can be shown \cite{abf} to behave like a Killing vector field
{\it on} the horizon, satisfying %
\begin{eqnarray}
l^a \nabla_a l^b~=~\kappa~ l^b~.
\label{ke}
\end{eqnarray}
Here, $\kappa$ is the acceleration of $l^a$ on the isolated
horizon. Unlike standard surface gravity whose normalization is fixed
by the requirement that the global timelike Killing vector generate
time translations at spatial infinity, the normalization of $\kappa$
here varies with rescaling of $l^a$.   

These features imply that while gravitational or
other radiation may exist arbitrarily close to the horizon, nothing
actually crosses the horizon, thereby emulating an `equilibrium'
situation. This, in turn, means that the area $A_H$ of the isolated
horizon must be a constant. Lifting of this restriction leads to
dynamical variants (the so-called `dynamical' horizons) which have
also been studied \cite{ak}; we shall however not consider these here. 

The actual implementation of these properties of the isolated horizon
require boundary conditions on the phase space variables on the
2-sphere foliate of the horizon. Recalling that the horizon is an inner
boundary of spacetime, it is obvious that one needs to add boundary
terms to the classical Einstein action, in order that the variational
principle can be used to derive equations of motion. It turns out
\cite{abk},\cite{abf} that the `boundary action' $S_H$
that one must add to the Einstein action (in the purely gravitational
case)
\begin{eqnarray}
S_E~=~{-i \over 8\pi G}~\int_{\cal M}~Tr \Sigma \wedge F~, 
\label{eha}
\end{eqnarray}
is an $SU(2)$ Chern Simons (CS) action\footnote{Strictly speaking, the
boundary conditions considered by \cite{abk} involve a partial gauge
fixing whereby the only independent connection on the horizon is
actually an internal radial $U(1)$ projection of the $SU(2)$ CS
connection. However, we ignore this subtlety at this point and
continue to work with the $SU(2)$ CS theory. The modification to the
final answer, had we chosen not to ignore this subtlety, will be
discussed later.} 
\begin{eqnarray}
S_H ~=~{-i \over 8\pi G}~{A_H \over 4\pi} \int_H Tr [ A \wedge dA ~+~
\frac23 ~A \wedge A \wedge A ] ~,
\label{cs}
\end{eqnarray}
where, now, $A$ is the CS connection, and $F$ the
corresponding curvature. The resultant modification to the symplectic
structure (\ref{sym}) is given by the CS symplectic two-form
\begin{eqnarray}
\Omega_H~=~-{k \over 2\pi}~\oint_S Tr[\delta ^{\gamma}A
\wedge \delta ^{\gamma}A']~,
\label{symcs}
\end{eqnarray}
where, $k \equiv A_H / 8\pi \gamma G$. In writing the boundary action 
(\ref{cs}), we have suppressed other terms like the boundary term
at infinity.

It is easy to see that the variational principle for the full action
is valid, provided we have, on the two-sphere foliation of $H$, the
restriction, 
\begin{eqnarray}
{k \over 2\pi}~F_{ab}^i~+~\Sigma_{ab}^i~=~0~.
\label{main}
\end{eqnarray}
Eq. (\ref{main}) has the physical interpretation of Gauss law for the
CS theory, with the two-form $\Sigma$ playing the role of source
current. We shall see shortly that this has crucial implications for
the quantum version of the theory.

\subsection{Quantum Aspects: General}
The classical configuration space consists of the space of smooth,
real $SU(2)$ Lie Algebra-valued connections modulo gauge
transformations \cite{almmt}. Alternatively, the space can be
described in terms of
three dimensional oriented, piecewise analytic networks or graphs 
embedded in the spatial slice $M$ \cite{almmt}.  Consider a particular
graph $C$ with $n$ links (or edges)
$e_1, \dots, e_n$; consider also the pullback of the connection $A$ to
$C$. Consider the holonomies defined as
\begin{eqnarray}
h_C(e_i) ~\equiv~{\cal P} \exp \oint_{e_i \in
C}  {^{\gamma}A_C}~, ~i=1,2 \dots, n~,
\label{hol}
\end{eqnarray}
where $^{\gamma}A_C$ represents the restriction of the connection to
the graph $C$; these span the configuration space ${\cal A}_C$ of
connections on the graph $C$. This space consists of $[SU(2)]^n$
group elements obtained as $n$-fold compositions of $SU(2)$ group
elements characterised by the spin $j_i$ of the edge $e_i$ for $i=1,2,
\dots, n$. The edges of $C$ terminate at vertices $v_1, \dots, v_n$
which, in their turn, are characterised by group elements $g(v_1),
\dots, g(v_m)$, which together constitute a set of $[SU(2)]^m$ group
elements for a given graph $C$. The union of spaces ${\cal A}_C$ for
all networks is then an equally good description of the classical
configuration space. 

The transition to the {\it quantum} configuration space is made,
first by enhancing the space of connections to include connections
$^{\gamma}{\bar A}$ which are not smooth but distributional, and then
considering the space ${\cal H}_C$ of square-integrable functions
$\Psi_C[^{\gamma}{\bar A}]$ of connections. For the integration
measure, one uses $n$-copies of the $SU(2)$-invariant Haar measure.
For a given network $C$, the wave function $\Psi_C[^{\gamma}{\bar
A}]$ can be expressed in terms of a smooth function $\psi$ of the
holonomies ${\bar h}_C(e_1), \dots, {\bar h}_C(e_n)$ of
distributional connections, 
\begin{eqnarray} \Psi_C[^{\gamma}{\bar
A}]~=~\psi({\bar h}_C(e_1), \dots, {\bar h}_C(e_n)) ~.  
\label{wf}
\end{eqnarray} 
The inner product of these wave functions can be defined as 
\begin{eqnarray} 
\langle \Psi_{1C}, \Psi_{2C}
\rangle~=~\int d\mu {\bar \psi}_{1C} \psi_{2C}~.  \label{inn}
\end{eqnarray} 
Basic dynamical variables include the holonomy operator ${\hat h}_C(e)$
and the operator version of the canonically conjugate
$\gamma$-rescaled  densitized triad $^{\gamma}{\hat E}_a^i$. The
holonomy operator acts diagonally on the wave functions,
\begin{eqnarray}
[{\hat h}_C(e) \Psi_C][^{\gamma}{\bar A}]~=~{\bar
h}_C(e)~\Psi_C[^{\gamma}{\bar A}] ~.  
\end{eqnarray}
The canonical conjugate densitized triad operators ${\hat E}_a^i$ act as
derivatives on $\Psi_C[^{\gamma} {\bar A}]$:
\begin{eqnarray}
^{\gamma}{\hat E}_a^i~\Psi[^{\gamma} A]~=~{\gamma~l_P^2 \over
i}~{\delta \over \delta ^{\gamma}A^a_i}~\Psi[^{\gamma}A]~.  
\end{eqnarray}
One defines the kinematical
Hilbert space ${\cal H}$ as the union of
the spaces of wave functions $\Psi_C$ for all networks.\footnote{
Unfortunately, ${\cal H}$ is {\it not} the  physical Hilbert space of
the theory; that space is the algebraic dual of ${\cal H}$ with no
natural scalar product defined on it. However, for the purpose of
calculation of the microcanonical entropy, it will turn out to be
adequate to use ${\cal H}$.} 

Particularly convenient bases for the wave functions are the spin
network bases. Typically, the spin network (spinet) states can be
schematically exhibited as 
\begin{eqnarray}
\psi_C(\{h_C\}; \{v\})~=~\sum_{\{m\}}~\prod_{v \in C}~I_v
~\prod_i~D^i_{\dots} ~,
\label{snet}
\end{eqnarray}
where, $D^i$ is the $SU(2)$ representation matrix corresponding to the
$i$th edge of the network $C$, carrying spin
$j_i$, and $I_v$ is the invariant $SU(2)$ tensor inserted at the
vertex $v$. If one considers all possible spin networks, the set of
spinet states corresponding to these is dense in the kinematical Hilbert
space ${\cal H}$. Spinet states diagonalize the densitized triad
(momentum) operators and hence operators corresponding to geometrical
observables like area, volume, etc. constructed out of the the triad
operators. The spectra of these observables turn out to be {\it
discrete}; e.g., for the area operator corresponding to the area of a
two dimensional spacelike physical surface $s$ (like the intersection of a
spatial slice with a black hole horizon), one considers the
spins $j_1,j_2, \dots, j_p$ on $s$ at the $p$ punctures made
by the $p$ edges of the spinet assumed to intersect the surface.
The area operator is defined as \cite{rs}, \cite{al}
\begin{eqnarray}
{\hat A_s}~ \equiv ~\left \{ \sqrt{n_a n_b{\hat E}^a_i{\hat E}^b_i}
\right \}_{reg}~ ,
\label{are}
\end{eqnarray}
where, $n_a$ is the normal to the surface, and ${reg}$ indicates
that the operator expression within the braces is suitably
regularized. The eigenspectrum turns out to be \cite{rs,al}
\begin{eqnarray}
a_s(p;\{j_i\})~=~8\pi \gamma l_P^2 \sum_{i=1}^p \sqrt{j_i(j_i+1)} ~.
\label{ares}
\end{eqnarray}
Spinet basis states correspond to networks without any `hanging' edge,
so that they transform as gauge singlets under the gauge group
$SU(2)$. Furthermore, invariance under spatial diffeomorphisms is
implemented by the stipulation that the length of any edge of any
graph is without physical significance. 

\subsection{Quantum Aspects: entropy calculation} 

As discussed in subsection A, the sphere $S_H$ formed by the
intersection of the isolated horizon and a spatial slice $M$ can be
thought of as an inner boundary of $M$. The dynamics of the isolated
horizon is described by an $SU(2)$ Chern Simons theory with the bulk
gravitational degrees of freedom playing the role of source
current. This picture can be implemented at the quantum level in a
straightforward manner. Because of the isolation implied by the
boundary conditions, the kinematical Hilbert space ${\cal H}$ can be
decomposed as
\begin{eqnarray}
{\cal H}~=~{\cal H}_V~\otimes~{\cal H}_S~, 
\end{eqnarray}
where, ${\cal H}_V~({\cal H}_S) $ corresponds to quantum states with
support on the spatial slice $M$ (on the inner boundary, i.e., these
are the Chern Simons states). The boundary conditions also imply the
Chern Simons Gauss law, eq. (\ref{main}); the quantum operator version
of this equation may be expressed as 
\begin{eqnarray}
{k \over 2\pi}~{\bf 1} \otimes  \hat{F}_{ab}^i~+~\hat {\Sigma}_{ab}^i
\otimes {\bf 1}~=~0~on ~S_H.
\label{maiq}
\end{eqnarray}
Now, the bulk spinet states diagonalize the operator $\hat{\Sigma}$
with distributional eigenvalues,
\begin{eqnarray}
\hat{\Sigma}(\vec{x})~|\psi \rangle_V \otimes | \psi
\rangle_S=~\gamma~  l_P^2 ~\sum_{i=1}^p
\lambda(j_i) \delta^{(2)} (\vec{x},\vec{x_i}) ~|\psi \rangle_V \otimes
| \psi \rangle_S .
\label{bul}
\end{eqnarray}
Eq. (\ref{maiq}) then requires that the boundary Chern Simons states
also diagonalize the Chern Simons curvature operator $\hat{F}$. In
other words, edges of the bulk spin network punctures the horizon foliate
$S_H$, endowing the $i$th puncture with a {\it deficit angle} \cite{abk}
$\theta_i~\equiv~\theta(j_i)$ for $i=1,2,\dots,p$, such that
\begin{eqnarray}
\sum_{i=1}^p \theta(j_i)~=~4\pi~.
\label{defi}
\end{eqnarray}
The curvature on $S_H$ is thus vanishingly small everywhere else
except at the
location of the punctures. This manner of
building up the curvature of the two-sphere $S_H$ out of a large but
finite number of deficit
angles requires that the number of such angles must be as large as
possible. This is achieved for the smallest possible value of all
spins $j_i$, namely $j_i=1/2 ~for~all~ i$. We shall come back to this
point later. 

The calculation of the entropy now proceeds by treating the isolated
horizon as a microcanonical ensemble with fixed area. Recalling the
semiclassical relationship between horizon area and mass of the
isolated horizon, this is equivalent to considering a standard
equilibrium microcanonical ensemble where the (average) energy of the
ensemble does not fluctuate thermally. The number of configurations of
such a system is equal to the exponential of the microcanonical
entropy $S_{MC}$. Likewise, in this case, the number of boundary Chern
Simons states $dim {\cal H}_S$ with pointlike sources, as depicted in
eq. (\ref{maiq}) (keeping (\ref{bul}) in view) yields $\exp
S_{MC}$. This number has been calculated for all four dimensional
non-rotating isolated horizons \cite{abk},\cite{km} of large macroscopic
fixed horizon area $A_H \gg l_P^2$. In ref. \cite{km}, the
computation makes use of the well-known relation between the
dimensionality of the boundary Chern Simons Hilbert space and the
number of {\it conformal blocks} of the corresponding two dimensional
$SU(2)_k$ Wess-Zumino-Witten model that `lives' on the punctured
two-sphere $S_H$. This number is given by
\begin{eqnarray}
dim~{\cal H}_S~=~\sum_p~\prod_{i=1}^p~\sum_{j_i} ~{\cal N}(p, \{j_i\})~,
\label{diml}
\end{eqnarray}
subject to the constraint that the area eigenvalues are fixed (to
within a fator of the Planck area) to the constant macroscopic area $A_H$,
\begin{eqnarray}
A_H~=~8\pi~ \gamma~ l_P^2~ \sum_{i=1}^p~ \sqrt{j_i(j_i+1)} ~,
\label{ares1}
\end{eqnarray}
where,
\begin{equation}
{\cal N}(p, \{j_i\})~=~\sum_{m_1= -j_1}^{j_1} \cdots
\sum_{m_p=-j_p}^{j_p} \left[
~\delta_{(\sum_{n=1}^p m_n), 0}~-~\frac12~ \delta_{(\sum_{n=1}^p m_n),
1}~-~\frac12 ~\delta_{(\sum_{n=1}^p m_n), -1} ~\right ]. \label{exct}
\end{equation}

Instead of the area constraint, one may now recall eq. (\ref{defi})
which also is a constraint on the spins and number of punctures. Using
this result in the area formula (\ref{ares1}) yields the maximal number
of punctures
\begin{eqnarray}
p_0~=~{A_H \over 4\pi~ \sqrt{3}~ \gamma~ l_P^2~}~ . 
\label{pmax}
\end{eqnarray}
The corresponding number of Chern Simons states for this assignment of
spins is given via (\ref{exct}) by
\begin{eqnarray}
{\cal N}(p_0)~\simeq~{2^{p_0} \over p_0^{\frac32}}~ \left[ 1 ~+~ const.~ +
~O(p_0^{-1})\right ]~.
\label{nmax}
\end{eqnarray}

Now, the (microcanonical) entropy of the isolated horizon is given by 
\begin{eqnarray}
S_{IH}~\equiv~\log~dim~{\cal H}_S~,
\end{eqnarray}
as remarked earlier. For isolated horizons with large macroscopic area
, the largest contribution to the {\it rhs} of eq.(\ref{diml}) is
given by the contribution of the single term of the multiple sum,
corresponding to $j_i=1/2 \forall i$ and $p=p_0$. This contribution
dominates all others in the multiple sum, so that, one has, using
eq.(\ref{nmax}), the microcanonical entropy formula \cite{km2}-\cite{pm}
\begin{eqnarray}
S_{IH}~=~S_{MC}~=~S_{BH}~+~\Delta_Q~,
\label{smc}
\end{eqnarray}
where, 
\begin{eqnarray}
S_{BH} ~\equiv~ A_H/4l_P^2
\end{eqnarray}
is the Bekenstein-Hawking Area Law (BHAL), and we have set the
Barbero-Immirzi parameter $\gamma=\log2/\pi \sqrt{3}$ \cite{abk} in
order to reproduce the BHAL with the correct normalization. $\Delta_Q$, 
given by
\begin{eqnarray}
\Delta_Q~=~-~\frac32 \log S_{BH}~+~const.~+~O(S_{BH}^{-1})~,
\label{qcor}
\end{eqnarray}
constitutes an infinite series (in decreasing powers of $S_{BH}$) of
corrections to the BHAL due to quantum fluctuations of spacetime, and
can be thought of as `finite size' corrections. One important aspect
of the formula (\ref{smc}) is that the coefficient of each correction
term is finite and unambiguously calculable, after $\gamma$ has been
fixed as mentioned.

\section{canonical entropy}
\subsection{Thermal fluctuations in a canonical ensemble}

In this subsection we present a derivation of the canonical entropy of
a standard equilibrium canonical ensemble, when small thermal
fluctuations around equilibrium are taken into account. The derivation
is a variant of the version given in \cite{dmb} and also in
textbooks. We begin with the formula for the canonical partition
function of a classical system in equilibrium 
\begin{eqnarray}
Z_C(\beta)~=~\int_0^{\infty}~dE~\exp-\beta E~\rho(E)~, 
\label{pf} 
\end{eqnarray}
where, $\rho(E)$ is the density of states. In what follows, we shall
employ the identification $\rho(E) \equiv \exp S_{MC}(E)$, where,
$S_{MC}(E)$ is the microcanonical entropy of an isolated subsystem
whose energy is held fixed at $E$. The integral in eq.(\ref{pf}) can
be performed in general by the saddle point approximation, provided
the microcanonical entropy $S_{MC}(E)$ can be Taylor-expanded around
the average equilibrium energy $E_0$,
\begin{eqnarray}
S_{MC}(E)~=~S_{MC}(E_0)~+~(E-E_0)~S'_{MC}(E_0)~+~
\frac12~(E-E_0)^2~S''_{MC}(E_0)~+~\dots~, 
\label{spa}
\end{eqnarray}
and higher order terms in powers of the energy fluctuation represented
by the $\dots$ in this expansion can be neglected in comparison to
terms of second order. The integration is then performed by requiring
that the inverse temperature $\beta$ is determined in terms of the
average energy $E_0$,
\begin{eqnarray}
\beta~=~S'_{MC}(E_0)~.
\label{bet}
\end{eqnarray}
The resulting Gaussian integration leads to the result 
\begin{equation}
Z_C(\beta)~=~e^{[-\beta E_0~+~S_{MC}(E_0)]}~\left[ {2\pi \over -
S''_{MC}(E_0)} \right]^{1/2} ~. 
\label{pfsp}
\end{equation}
Using the standard formula from equilibrium statisticl mechanics, 
\begin{eqnarray}
S_C~=~\beta~E_0~+~\log Z_C
\end{eqnarray}
it is easy to deduce that the canonical entropy is given in terms of
the microcanonical entropy by
\begin{eqnarray}
S_C(E_0)~=~S_{MC}(E_0)~+~\Delta_F~,
\label{scan}
\end{eqnarray}
where,
\begin{eqnarray}
\Delta_F~\equiv~\frac12 \log[-S''_{MC}(E_0)] ~,
\label{fcor}
\end{eqnarray}
is the leading correction to the canonical entropy due to thermal
fluctuations in the energy. Using the definition of the specific heat
of the system
\begin{eqnarray}
C~\equiv~-\beta^2~\left({\partial E_0 \over \partial \beta} \right)~, 
\label{spht}
\end{eqnarray}
this correction may be reexpressed as
\begin{eqnarray}
\Delta_F~=~\frac12 \log ~\left({C \over \beta^2}\right)~. 
\label{fl}
\end{eqnarray}
Clearly formulae (\ref{fcor})-(\ref{fl}) make sense provided
$-S''_{MC} < 0$, i.e., the specific heat is positive, as is true in
general for equilibrium thermodynamic systems. 

It is interesting that thermal fluctuations produce a positive
correction to the canonical entropy. This is in contrast to the case
of the black hole microcanonical entropy where QGR effects tend to
reduce the semiclassical value by restricting the set of degenerate
states to those that are singlets under the residual gauge group
$SU(2)$.

\subsection{Canonical entropy of anti de Sitter black holes} 

The corrections to the canonical entropy due to thermal fluctuations
can be calculated in principle for all isolated horizons which
includes all stationary black holes. We shall first deal with the case
of adS black holes where the calculation makes sense for a certain
range of parameters of the black hole solution. Computation of such
corrections has been performed in ref. \cite{dmb}. Here, we recount
the computation in a slightly different form, and compare the result
with the corrections to the BHAL due to quantum spacetime
fluctuations. 

\subsubsection{BTZ}

The non-rotating BTZ metric is given by \cite{btz}
\begin{eqnarray}
ds^2~=~-~\left({r^2 \over \ell^2} ~-~8G_3M \right)~dt^2~+~\left ({r^2
\over \ell^2} ~-~8G_3M \right)^{-1}~dr^2 ~+~r^2~d\phi^2~,
\label{btz}
\end{eqnarray}
where, $\ell^2 \equiv -1/\Lambda^2$ and $\Lambda$ is the cosmological
constant. The BH entropy is
\begin{eqnarray}
S_{BH}~=~{\pi r_H \over 2G_3}~,
\end{eqnarray}
where, the horizon radius $r_H = \sqrt{8G_3 M \ell}$. Quantum
spacetime fluctuations produce corrections to the
microcanonical BHAL, given for $r_H \gg \ell$  by \cite{gks}
\begin{eqnarray}
\Delta_Q~=~-~\frac32 ~\log S_{BH}~. 
\label{btzq}
\end{eqnarray}
Using (\ref{btzq}), and identifying the mass $M$ of the black hole
with the equilibrium energy $E_0$, the microcanonical entropy $S_{MC}$
has the properties
\begin{eqnarray}
S''_{MC}(M)~& < &~0~ for~ r_H > \ell \nonumber \cr
S''_{MC}(M)~& > &~0~ for ~r_H < \ell ~.
\label{smcf}
\end{eqnarray}
Alternatively, the specific heat of the BTZ black hole is positive, so
long as $r_H \geq \ell$. The system can therefore be thought of as being
in equilibrium for parameters in this range. It follows that the
calculation of $\Delta_F$ yields a sensible result in this range,
\begin{eqnarray}
\Delta_F~=~\frac32~ \log S_{BH}~=~-~\Delta_Q~.
\label{btzf}
\end{eqnarray}
The import of this for the canonical entropy is rather intriguing,
using eq.(\ref{scan}) 
\begin{eqnarray}
S_C~=~S_{BH}~.
\label{nonr}
\end{eqnarray}
The quantum corrections to the BHAL in this case are cancelled by
corrections due to thermal fluctuations of the area (mass) of the
black hole horizon. We do not know the complete significance of this
result yet.\footnote{We should mention that this result ensues only if
one takes recourse to the classical relation between the horizon area
and the mass. The validity of that relation in the domain in which the
QGR calculation has been performed, is not
obvious at this point.}

\subsubsection{4 dimensional adS Schwarzschild}

Such black holes have the metric
\begin{eqnarray}
ds^2~=~-V(r)~dt^2~+~V(r)^{-1}~dr^2~+~r^2~d\Omega^2~, 
\label{ads4}
\end{eqnarray}
where, 
\begin{eqnarray}
V(r) ~=~1~-~{2GM \over r}~+~{r^2 \over \ell^2} ~,
\label{pot}
\end{eqnarray}
with $\ell^2 \equiv -3/ \Lambda$. The horizon area $A_H = 4\pi r_H^2$,
where the Schwarzschild radius obeys the cubic $V(r_H) =0$. It is easy
to see that the cubic yields the mass-area relation
\begin{eqnarray}
M~=~\frac{1}{2G}~\left({A_H \over 4\pi} \right)^{1/2} ~\left(1~+~{A_H
\over 4\pi \ell^2} \right)
\label{mar}
\end{eqnarray}
It is clear from eq.(\ref{mar}) that 
\begin{eqnarray}
S''_{MC}(M)~<~0~for~A_H~>~\frac43 \pi \ell^2~,
\label{adsr}
\end{eqnarray}
so that, once again the specific heat is positive in this range. The
thermal fluctuation contribution for this parameter range is
\begin{eqnarray}
\Delta_F~=~\log \left({A_H \over l_P^2} \right) ~.
\label{adsf}
\end{eqnarray}
The net effect on the canonical entropy is a partial cancellation of
the effects due to quantum spacetime fluctuations and thermal
fluctuations,
\begin{eqnarray}
S_C~=~S_{BH}~-~\frac12~\log S_{BH}~.
\label{adsc}
\end{eqnarray}
Note that the thermal and quantum fluctuation effects compete with
each other in both cases considered above, with the net result that
the canonical entropy is still superadditive
\begin{eqnarray}
S_C(A_1~+~A_2) ~\geq~S_C(A_1)~+~S_C(A_2) ~.
\end{eqnarray}

The point $r_H \sim \ell$ in parameter space signifies the breakdown
of thermal equilibrium; this point has been identified with the
so-called Hawking-Page phase transition \cite{hp} from the black hole
phase to a phase which has been called an `adS gas'. In this latter
phase, the black hole is supposed to have `evaporated away', leaving
behind a gas of massless particles in an asymptotically adS spacetime.

\section{4 dimensional asymptotically flat black holes} 

One may want to repeat the computation of the thermal fluctuation
correction to the canonical entropy for black holes which are
asymptotically flat, i.e., the Kerr-Newman family of solutions of the
Einstein-Maxwell equations without a cosmological constant. However,
this is stymied by a pathology which can be seen most easily from the
case of the Schwarzschild black hole: the classical relation between
horizon area $A_H$ and mass $M$ is the well known $A_H = 16\pi G
M^2$. If one uses this relation in the formula for the microcanonical
entropy (\ref{smc}), it is obvious that $S''_{MC}(M) > 0$ for all
nonvanishing positive values of the mass. The canonical entropy, as a
consequence, acquires an imaginary part, signifying a thermodynamic
instability. Alternatively, the specific heat turns out to be
negative. Thus the standard approach of including the effet of thermal
fluctuations around equilibrium fails in this case, as the canonical
description is no longer adequate. This failure persists with the
inclusion of electric charge and angular momentum, so long as one has
a well-defined stationary event horizon and stays away from
extremality. It is therefore a generic conundrum vis-a-vis the use of the
canonical ensemble in such cases. The only way asymptotically flat
black holes can be described in terms of equilibrium ensembles is by
restricting the energy available to them to a constant, i.e., by using
a microcanonical ensemble. Unphysical as it may be, there does not
appear to be any alternative, if we continue to use the {\it classical}
relation between black hole mass and horizon area.  

The origin of the malaise can be traced \cite{hp} to the
extraordinarily large degeneracy of asymptotically flat black holes,
delineated in the density of states growing as $\rho(M) \sim \exp
M^2$, notwithstanding the power law suppression due to quantum
spacetime fluctuations. Defining the classical canonical partition
function as
\begin{eqnarray}
Z_C(\beta) ~=~\int_0^{\infty} dM~\exp -\beta M~\rho(M)~ ,
\label{bhpf}
\end{eqnarray}
it is obvious that the integral can never converge for large
$M$. Contrast this to the case of the adS Schwarzschild black hole for
horizon areas $A_H > \frac43 \pi \ell^2$; the density of states grows
only as $\rho(M) \sim \exp M^{2/3}$ for large masses, allowing thereby the
Boltzmann factor to tame the integral in (\ref{bhpf}). 

As we had remarked earlier (footnote 4), the validity of the classical
relation between mass and horizon area, and identifying the mass with
the internal energy, are of course not guaranteed in the domain of
QGR. We start with the canonical partition
function in the quantum case
\begin{eqnarray}
Z_C(\beta)~=~Tr~\exp -\beta \hat{H} ~.
\label{qpf}
\end{eqnarray}
Recall that in classical general relativity in the Hamiltonian
formulation, the bulk Hamiltonian is a first class constraint, so that
the entire Hamiltonian consists of the boundary contribution $H_S$
on the constraint surface. In the quantum domain, the Hamiltonian
operator can be written as
\begin{eqnarray}
\hat{H}~=~\hat{H}_V~+~\hat{H}_S~,
\end{eqnarray}
with the subscripts $V$ and $S$ signifying bulk and boundary terms
respectively. The Hamiltonian constraint is then implemented by requiring
\begin{eqnarray}
\hat{H}_V~|\psi \rangle_V~=~0~
\label{hcon}
\end{eqnarray}
for every physical state $|\psi \rangle_V$ in the bulk Hilbert space.
This relation implies that the partition function may be written as 
\begin{eqnarray}
Z_C~=~\sum_{V,S}~_S{\langle \chi |\exp -\beta \hat{H}_S
|\chi \rangle}_S~.
\label{rpf}
\end{eqnarray}
Thus, the relevance of the bulk physics seems rather limited due to
the constraint (\ref{hcon}). The partition function thus reduces to
\begin{eqnarray}
Z_C(\beta)~=~dim~{\cal H}_V~Z_S(\beta)~,
\label{redp}
\end{eqnarray}
where $Z_S$ is the `boundary' partition function given by
\begin{eqnarray}
Z_S(\beta)~=~Tr_S~\exp -\beta \hat{H}_S~.
\label{zs}
\end{eqnarray}
Since we are considering situations where, in addition to the boundary
at asymptopia, there is also an inner boundary at the black hole
horizon, we may infer that quantum fluctuations of this boundary lead
to black hole thermodynamics. The factorization in eq.(\ref{redp})
manifests in the canonical entropy as the appearance of an additive
constant proportional to $dim~{\cal H}_V$. Since thermodynamic entropy
is defined only upto an additive constant, perhaps one may argue that
the bulk states do not play any role in black hole thermodynamics.  It
is not clear yet if this can be thought of as the origin of the
holographic hypothesis \cite{thf}. 

The next step is to evaluate the boundary partition function
$Z_S$. For this, one needs to express the boundary Hamiltonian
$\hat{H}_S$ as a function of other observables pertaining to the
boundary, notably the area operator. Since the horizon states $|\chi
\rangle_S$ are an eigenbasis for the area operator, the boundary
partition function can be written as
\begin{eqnarray}
Z_S(\beta)~=~\sum_p~\sum_{j_1}~\cdots~\sum_{j_p}~ [\exp -\beta {\cal
E}(a_S(p;\{j_k\}))]~{\cal N}(p;\{j_k\}) ~,   
\label{spf}
\end{eqnarray}
where, $a_S$ is given by eq..(\ref{ares}), and ${\cal N}(p;\{j_l\})$
by eq.(\ref{exct}), and the function ${\cal E}(a_S)$ is to be
determined, after appropriate regularization, in QGR. It is not
difficult to show that lifting the classical relation between black
hole mass and area to the quantum level na\"ively, for instance for
the ordinary Schwarzschild black hole, does not lead to a convergent
$Z_S$. The sum over $p$ in eq.(\ref{spf}) invariably diverges for
large $p$, once again because of the exponentially large contribution
from the degeneracy factor ${\cal N}$ in the partition function. In
the Appendix, we sketch a derivation of this result for the case when
all spins are indentical. This means that even within QGR the
thermodynamic instability discerned in the classical analysis persists,
if we na\"ively use classical relations at the operator level. We
mention in passing that there is a possible {\it lower} bound on the
area eigenvalue spectrum. It is clear that a physical 2-surface $S$
must have at least two punctures with spins $j_1=j_2=1/2$ to
correspond to an $SU(2)$ singlet state. This leads to an area
eigenvalue
\begin{eqnarray}
a_S^{min}~=~8 l_P^2 \log2 ~, with~\gamma=\log2/\pi \sqrt{3}~.
\label{amin}
\end{eqnarray}
While we do not know at this point the actual physical significance of
this minimal area, it is amusing to speculate that black hole
radiation culminates in a remnant of this size. 

\section{Conclusions}

The exact cancellation of the logarithmic corrections to the BHAL for
the BTZ black hole canonical entropy, due to quantum spacetime
fluctuations and thermal fluctuations, may hold a deeper significance
which warrants further analysis. In ref. \cite{gks}, the
microcanonical entropy is calculated from the exact Euclidian
partition function of the $SU(2) \times SU(2)$ Chern Simons theory
which describes the BTZ black hole. Perhaps this calculation can be
extended to a finite temperature canonical quantum treatment of the
problem, to include the effect of thermal fluctuations. Such an
analysis is necessary to allay suspicions about using (semi)classical
relations between the mass and the area of the black hole. Likewise,
for four dimensional black holes, the precise relation admitted within
QGR between the boundary Hamiltonian and the area operator needs to be
ascertained. Presumably, this relation will not qualitatively change
the results for adS black holes, although it might lead to a better
understanding of the Hawking-Page phase transition. More importantly,
this issue needs to be addressed in order to determine if the
thermodynamic instability found for generic asymptotically flat black
holes is an artifact of a semiclassical approach.

It is conceiveable that inclusion of charge and/or angular momentum
for adS black holes in dimensions $\geq 4$ will present no conceptual
subtleties, so long as the (outer) horizon area exceeds in magnitude
the inverse cosmological constant. However, to the best of our
knowledge, the formulation of a higher dimensional (i.e., $>$ 4) QGR
has not been completed; this will have to be done before comparison of
quantum and thermal fluctuation effects can be made in higher
dimensions. 

The thermodynamic instability discerned for asymptotically flat black
holes also appears to emerge for de Sitter Schwarzschild black holes
\cite{dmb}. Since current observations appear to point enticingly to
an asymptotically de Sitter universe, this instability must be better
understood. In its present incarnation, it would imply that massive
black holes will continue to get heavier {\it without limit}. On the
other end of the scale, the instability can be interpreted in terms of
disappearance of primordial black holes due to Hawking radiation,
except for the possible existence of Planck scale remnants. It should
be possible to estimate the density of such remnants and check with
existing bounds from cosmological data.

One of us (PM) thanks A. Ashtekar, R. Bhaduri, S. Das, G. Date,
T. R. Govindarajan, R. Kaul, J. Samuel, S. Sinha and G.'t Hooft for 
useful discussions.

\section{Appendix}

We begin with the formula (\ref{spf}) for the boundary partition function
\begin{eqnarray}
Z_S(\beta)~=~\sum_p~\sum_{j_1}~\cdots~\sum_{j_p}~ [\exp -\beta {\cal
E}(a_S(p;\{j_k\}))]~{\cal N}(p;\{j_k\}) ~.
\label{spf}
\end{eqnarray}
For the ordinary Schwarzschild black hole, classically the black hole
mass $M=(A_H/16\pi G)^{1/2}$; the quantum generalization of this can
be taken to be
\begin{eqnarray}
{\cal E}(a_S(p;\{j_i\})~=~\left({a_S(p;\{j_i\} \over 16\pi G}\right)^{1/2}
\label{mare}
\end{eqnarray}
Now, for simplicity, let all spins $j_i = J ~\forall~ i=1,2,\dots,n$;
the multiple sum over the spins then reduce to a single one over all
possible half-integral values of $J$
\begin{eqnarray}
Z_S(\beta)~=~\sum_p~\sum_J~[\exp -\beta (a_S(p;J)/16\pi G)^{1/2}]~{\cal
N}(p;J)  ~,
\label{spfj}
\end{eqnarray}
where, 
\begin{eqnarray}
a_S(p;J)~=~8\pi \gamma l_P^2 p\sqrt{J(J+1)} ~,
\label{asj}
\end{eqnarray}
and
\begin{eqnarray}
{\cal N}(p;J)~\simeq~{(2J+1)^p \over p^{f(J)}}~for~large~p~.
\label{npj}
\end{eqnarray}
The function $f(J)$ is at most a polynomial for large $J$. 

The object is to argue that the summation over $p$ in eq.(\ref{spfj})
diverges for large $p \gg 1$. This is best demonstrated by replacing
the double sum in (\ref{spfj}) with double integrals over $p$ and $J$
and using (\ref{asj}) and (\ref{npj}), 
\begin{eqnarray}
Z_S(\beta)~\sim~\int^{\infty}_{p_0}
dp~\int^{\infty}_{J_0}~{1 \over p^{f(J)}}~\exp~[-~Kp^{1/2}(J(J+1))^{1/4}
~+~ p \log(2J+1)],
\label{inpj}
\end{eqnarray}
where, the dimensionless constant $K \sim \gamma \beta l_P$. It is
obvious from eq.(\ref{inpj}) that for large $p$, the $O(p)$ term in
the exponent dominates the integral, irrespective of how large $J$ is,
and returns to the same divergent behaviour as in the classical case.

\end{document}